\documentclass[12pt,fleqn]{article}
\usepackage[T1]{fontenc}
\usepackage{geometry}
\usepackage{palatino}
\usepackage{mathpazo}
\usepackage{mathtools}
\usepackage{amssymb} 
\usepackage[nolist]{acronym}
\usepackage{dsfont} 
\usepackage[small]{caption} 
\usepackage{mathrsfs}	
\usepackage{cite} 
\usepackage{cancel}
\usepackage{nicefrac}
\usepackage{pifont}
\usepackage{tikz}
\definecolor{NavyBlue}{rgb}{0.0, 0.0, 0.5}
\usepackage[linkcolor=blue,urlcolor=blue,citecolor=NavyBlue,colorlinks=true]{hyperref} 
\usepackage{eqparbox}
\usepackage[smalltableaux]{ytableau}
\usepackage{young}
\usepackage{braket}
\usepackage{xfrac}
\usepackage{xspace}
\usepackage{booktabs}
\usepackage{mleftright}
\mleftright
\usepackage{subcaption}
\usepackage{letltxmacro}

\usepackage[textsize=tiny]{todonotes}
\usepackage{soul}
\usepackage{cleveref} 

\makeatletter
\g@addto@macro\bfseries{\boldmath}
\makeatother

\DeclareMathOperator{\Pf}{Pf}

\newcommand{\E}[1]{\ensuremath{\mathrm{E}_{#1}}}

\newcommand{\SU}[1]{\ensuremath{\mathrm{SU}(#1)}}
\newcommand{\SP}[1]{\ensuremath{\mathrm{Sp}(#1)}}

\newcommand{\U}[1]{\ensuremath{\mathrm{U}(#1)}}
\newcommand{\Z}[1]{\ensuremath{\mathds{Z}_{#1}}} 
\newcommand{\x}{\ifmmode\times\else\ensuremath{\times}\fi}

\newcommand{\rep}[1]{\ensuremath{{\boldsymbol{#1}}}}
\newcommand{\crep}[1]{\ensuremath{\overline{\boldsymbol{#1}}}}

\newcommand{\FTT}{\ensuremath{4{\leadsto}3}\xspace}
\newcommand{\TTT}{\ensuremath{2{\leadsto}3}\xspace}

\newcommand{\SCONF}{$s$-confinement\xspace}
\newcommand{\SCONFG}{$s$-confining\xspace}
\newcommand{\SUTS}{\ifmmode\SU2_s\else$\SU2_s$\xspace\fi}
\LetLtxMacro\OldDot\.
\renewcommand{\.}{\ifmmode\hspace{0.1ex}\else\OldDot\fi}
\LetLtxMacro\OldOverline\overline
\renewcommand{\overline}[1]{{\hspace{0.1ex}\OldOverline{\hspace{-0.1ex}#1\hspace{-0.033ex}}\hspace{0.033ex}}}

\let\Composite\mathcal

\allowdisplaybreaks

\begin{acronym}
\acro{GUT}{Grand Unified Theory}
\acro{IR}{infrared}
\acro{NS}{Nelson and Strassler}
\acro{RG}{renormalization group}
\acro{NS}{Nelson--Strassler}
\acro{RT}{Razamat--Tong}
\acro{SM}{Standard Model}
\acro{MSSM}{minimal supersymmetric Standard Model}
\acro{SYM}{super-Yang--Mills}
\acro{VEV}{vacuum expectation value}
\acro{UV}{ultraviolet}
\end{acronym}

\begin{document}
\newcommand\mytitle{Generation flow in field theory and strings}
\newcommand\mypreprint{UCI-TR-2021-21}
\begin{titlepage}
\begin{flushright}
\mypreprint
\end{flushright}

\vspace*{2cm}

\begin{center}
{\Large\sffamily\bfseries\mytitle}

\vspace{1cm}

\renewcommand*{\thefootnote}{\fnsymbol{footnote}}

\textbf{%
Sa\'ul~Ramos--S\'anchez$^{a,}$\footnote{ramos@fisica.unam.mx}, 
Michael~Ratz$^{b,}$\footnote{mratz@uci.edu},
Yuri~Shirman$^{b,}$\footnote{yshirman@uci.edu},
Shreya~Shukla$^{b,}$\footnote{sshukla4@uci.edu} and
Michael~Waterbury$^{b,}$\footnote{mwaterbu@uci.edu}}
\\[8mm]
\textit{$^a$\small Instituto de F\'isica, Universidad Nacional Aut\'onoma de M\'exico, POB 20-364, Cd.Mx. 01000, M\'exico}
\\[5mm]
\textit{$^b$\small
~Department of Physics and Astronomy, University of California, Irvine, CA 92697-4575 USA
}
\end{center}

\vspace*{1cm}
\begin{abstract}
Nontrivial strong dynamics often leads to the appearance of chiral composites.
In phenomenological applications, these can either play the role of Standard Model 
particles or lift chiral exotics by partnering with them in mass terms. 
As a consequence, the RG flow may change the effective number of chiral 
generations, a phenomenon we call generation flow.  We provide explicit 
constructions of globally consistent string models exhibiting generation flow. 
Since such constructions were misclassified in the traditional model searches, 
our results imply that more care than usually appreciated has to be taken when 
scanning string compactifications for realistic models.
\end{abstract}
\vspace*{1cm}
\end{titlepage}

\section{Introduction}
\label{sec:Introduction}

One of the curious features of the \ac{SM} of particle physics is the repetition
of families. That is, the matter content of the \ac{SM} comprises three copies
of fermions carrying identical \ac{SM} gauge quantum numbers. While the number
of generations is generally arbitrary in field theoretic extensions of the
\ac{SM}, such as a \ac{GUT}, in string theory it can be thought of as a
prediction of any specific model or compactification. Hence, the number of
generations is often used as one of the first selection filters applied in a
search for promising string models. It is the purpose of this study to point
out that non-perturbative field theoretic dynamics may modify the number of
effective generations in the process of \ac{RG} flow. Thus, some additional
care is required when counting the number of generations in candidates for
\ac{UV} completions of the \ac{SM}, in particular in string models.

In this paper, we will concentrate on supersymmetric models both because it is 
convenient in the context of string model building and because the relevant 
non-perturbative dynamics is under qualitative and often quantitative control
in  such theories. As shown by Seiberg~\cite{Seiberg:1994bz}, non-perturbative
effects can have a dramatic impact on gauge theories. In particular, due to
confinement  and duality, the degrees of freedom appropriate for describing
\ac{IR} physics  often differ considerably from the \ac{UV} degrees of freedom.
Throughout this paper, aiming at preserving the chirality of the \ac{SM}  (or
its \ac{GUT} completion),  we consider confinement without chiral symmetry
breaking (so-called  $s$-confinement~\cite{Csaki:1996sm,Csaki:1996zb}). Since
the low-energy degrees of freedom in these models are composites of the 
elementary fields, they usually transform in different representations of the 
unbroken global symmetry. 
When a subgroup of such global symmetry is identified with a GUT or the \ac{SM} gauge group, 
a new, composite, chiral generation may emerge in the \ac{IR} or, alternatively,
an existing chiral generation may become massive. The first of these phenomena
was initially used in ~\cite{Strassler:1995ia, Nelson:1996km} to construct realistic
extensions of the minimal supersymmetric Standard Model with some of the third generation quarks and Higgs
bosons arising as composites of strong dynamics. In this approach, which we will
refer to as the \ac{NS} mechanism, the \ac{RG} flow leads to the appearance of
light chiral composites in the \ac{IR} thus increasing the effective number of
chiral generations. The NS mechanism may be modified in several fairly obvious
ways. For example, some of the composites may acquire masses by
mixing with elementary chiral fields, modifying the spectrum of light fields in
the \ac{IR} in nontrivial ways.  When all of the composites acquire mass, the model is in the second regime which attracted attention more recently \cite{Razamat:2020kyf}. We will refer to the second phenomenon as the \ac{RT} mechanism.
Here all of the composites  of strong dynamics acquire masses by partnering with
elementary degrees of freedom  and thus reduce the number of effective
generations in the \ac{IR}. As we will argue,  these two mechanisms can be
continuously connected by introducing mass terms for  vector-like elementary
fields, which are allowed to mix with the composites. When  masses of
vector-like fields are small while the mixing between elementary fields  and
composites is of order one, the theory flows to the \ac{RT} limit where all  the
light fields are elementary. On the other hand, in the limit of large mass  the
vector-like elementary fields decouple, leaving massless composites behind.  In
this case, the theory flows to the \ac{NS} limit where some light fields are 
composites. By varying the mass terms, one can interpolate between the two
limits, and for intermediate values of the mass term some \ac{IR} degrees of
freedom will be  partially composite. Furthermore, one has freedom to decouple any
number of composites. In general, however, non-perturbative dynamics affects
\ac{RG} flow  and modifies the effective number of chiral generations in the
\ac{IR}. We will refer to these phenomena as \emph{generation flow}. 

It is then natural to ask whether generation flow can occur in scenarios where 
the number of generations is predicted from other data. This is particularly 
relevant for string model building (cf.\ e.g.\ \cite{Ibanez:2012zz} for a review), 
where one obtains the \ac{SM} generations from string compactifications. We will 
argue that generation flow indeed occurs in some globally consistent string models.
In these constructions, the true number of generations in the \ac{IR} description 
can differ from the tree-level value that one obtains at the compactification
scale.
Hence, a search for 3-generation models in string theory has to
go beyond the tree-level analysis.

\enlargethispage{\baselineskip}
This paper is organized as follows. In \Cref{sec:SU6model}, we will review the 
\ac{RT} mechanism of gapped chiral fermions. In \Cref{sec:SU5_language}, we 
construct  models exhibiting generation flow towards a 3-generation theory with 
(a \ac{GUT} completion of) the \ac{SM} gauge group in the \ac{IR}. Our first 
example is a \FTT model based on the \ac{RT} mechanism where all the \ac{IR}
degrees  of freedom are elementary. We then construct a generalization of the
\FTT model  where some of the third generation fields are composite. We point
out that our  construction is analogous to the \ac{NS}
mechanism~\cite{Strassler:1995ia,Nelson:1996km}. This  motivates us to build a
\TTT model with an upward generation flow. Furthermore, we discuss the stability
of the chirally symmetric vacua in $s$-confining models under the  deformations 
which induce generation flow. While such deformations may generally  destabilize 
the vacua by non-perturbative dynamics (see~\cite{RSSW} for a more  detailed discussion), 
we argue that the chirally symmetric vacua survive in our models. In
\Cref{sec:Explicit_string_models}, we collect evidence for the existence of 
string models exhibiting generation flow by presenting explicit examples.
Finally, \Cref{sec:Summary} contains our conclusions.

\section{\SCONF and gapped chiral fermions}
\label{sec:SU6model}

We begin by briefly reviewing dynamics of supersymmetric gapped fermion models 
introduced in \cite{Razamat:2020kyf}. In the following we will take the approach
of \cite{RSSW} to building models of chiral gapped fermions. This approach
starts with SUSY QCD models  that exhibit confinement without chiral symmetry
breaking on smooth moduli space \cite{Seiberg:1994bz}.\footnote{This dynamics is
usually referred to as \SCONF. See \cite{Csaki:1996zb} for a complete
classification of such theories.}  For our purposes it is convenient to restrict
attention to \SCONF in \SUTS SUSY QCD with six chiral doublet superfields and
thus $\SU6$ chiral global symmetry. We review the dynamics of this model in the
subsection \ref{sec:sconfine}. In the subsection \ref{sec:gap}, we discuss the
deformation of the SUSY QCD required to arrive at mass gap models of
\cite{Razamat:2020kyf}.

\subsection[s-confining SU(2)\_s model]{\SCONFG \SUTS model} 
\label{sec:sconfine}

The model outlined above possesses $\SU6\times\SUTS$ symmetry, where $\SU6$ is a
chiral global symmetry while \SUTS is a strongly interacting \SCONFG gauge
group. For future convenience we will assign quark superfields to
$(\crep{6},\rep{2})$ representation of the symmetry group. The theory possesses
a set of classical $D$-flat directions which can be parameterized either in
terms of squark \acp{VEV} or in terms of gauge invariant mesons which are
classically defined as $M_{ij}\sim Q_iQ_j/\Lambda$, where we suppressed
contraction of \SUTS-color indices and the dynamical scale of the quantum theory
$\Lambda$ is introduced on dimensional grounds. The mesons $M$ transform in the
conjugate antisymmetric representation  of the global $\SU6$ symmetry
$\crep{15}$. However, since quark \acp{VEV} satisfy a set of
algebraic identities, not all meson \acp{VEV} are independent. These classical
constraints imply a set of relations between the mesons,
\begin{equation}
 \varepsilon^{i_1\ldots i_6}M_{i_3i_4}M_{i_5i_6}=0\,.
\end{equation}
One may implement these constraints in the composite description of the theory
by postulating a dynamical superpotential
\begin{equation}
\label{eq:sconfine}
 \mathscr{W}_s = \varepsilon^{i_1\ldots i_6}M_{i_1i_2}M_{i_3i_4}M_{i_5i_6}\equiv \Pf(M)\,.
\end{equation}
The moduli space parameterized by mesons $M$ together with the
superpotential~\eqref{eq:sconfine}  coincides with the classical moduli space of
the theory parameterized by quark \acp{VEV} satisfying $D$-flatness conditions. It
was shown in \cite{Seiberg:1994bz} that the classical moduli space of vacua remains
unmodified quantum mechanically and the IR physics is described in terms of
weakly interacting mesons with the superpotential~\eqref{eq:sconfine}. While the
chiral global symmetry of this model is broken at a generic point on the moduli
space, the chiral symmetry remains unbroken at the origin where the theory
exhibits confinement without chiral symmetry breaking. This is precisely the
vacuum we are interested in.

\subsection{Mass gap model}
\label{sec:gap}

For phenomenological purposes we are interested in gauging $\SU6$ global
symmetry of the \SCONFG model discussed in the previous subsection (more
precisely we are interested in gauging a subgroup of $\SU6$, such as a GUT 
$\SU5$ or the \ac{SM} group $\SU3\times\SU2\times\U1$). To this end, 
one must introduce a set of spectator fields charged under $\SU6$ but not 
\SUTS (so that the \SCONFG dynamics remains unaffected) to ensure a 
cancellation of the cubic $\SU6$ anomaly. This can be achieved, for example, 
by introducing spectators that transform in representations of
$\SU6$ conjugate to those of elementary fields, i.e.\ by adding two spectators with
quantum numbers given by $(\rep{6},\rep{1})$. Alternatively, one can introduce a
single spectator $S$ in an \SU6 representation conjugate to the one of the
mesons, i.e.\ transforming as $(\rep{15},\rep{1})$. In the former case, the
theory remains chiral both in the \ac{UV} and \ac{IR}.  This is because \SUTS is
not yet confined in the \ac{UV} and the matter fields transform in chiral
representations of the full $\SU6\times \SUTS$ symmetry, while the
representations of \ac{IR} degrees of freedom are chiral under \SU6. However, in
the latter case, the chiral properties of the model change as the theory flows
from the \ac{UV} to the \ac{IR}. While the \ac{UV} theory is clearly chiral, the
\ac{IR} degrees of freedom, the mesons $M$ and spectators $S$, transform in
conjugate representations and thus form a single vector-like
representation. By choosing to cancel anomalies with the spectator $S$ in the
antisymmetric representation, we will be able to construct a model  flows
from a gapless, chiral phase in the \ac{UV} to a gapped phase in the \ac{IR}.

Since the matter content in the \ac{IR} is non-chiral, a mass term, $SM$,  is
allowed in the \ac{IR} superpotential. In terms of the \ac{UV} degrees of
freedom, this mass term corresponds to a marginal operator, $SQ^2$. Thus, we
deform the \SCONFG model by a tree-level superpotential
\begin{equation}
\label{eq:su2s-tree}
 \mathscr{W}=y\.S\.Q^2=c\.\Lambda\. S\.M\,,
\end{equation}
where the numerical coefficient $c$ represents both an arbitrary Yukawa coupling
$y$ of the \ac{UV} theory and the fact that the mass scale generated by
confinement is not directly calculable. 

At this point one might be tempted to conclude that a mass gap develops in the
chirally symmetric vacuum at the origin, while the rest of the  moduli space is
lifted by the equations of motion for $S$ and $M$. However, while ultimately correct,
this conclusion is somewhat premature. Indeed, while lifting $\SUTS$ D-flat directions, the deformation \eqref{eq:su2s-tree}
introduces new classical flat directions, those parameterized by \SUTS singlets
$S$. Since any \acp{VEV} for $S$ would break the chiral symmetry, it is
important  to verify that the non-perturbative
dynamical superpotential \eqref{eq:sconfine} does not destabilize these
directions. A careful analysis \cite{RSSW} of the full superpotential in
\eqref{eq:sconfine} and \eqref{eq:su2s-tree} demonstrates that \SUTS dynamics
generates an effective superpotential for gauge singlets $S$ stabilizing them at
the origin.\footnote{We stress that this conclusion is model dependent, and
there exist models where the $S=0$ vacuum at the origin is destabilized,
resulting in chiral symmetry breaking.} While referring the reader to \cite{RSSW}
for the full analysis, we present a simple argument here. Consider the theory at
large $S$ where all quark superfields become heavy. In this region of the moduli
space the low-energy physics is described in terms of a pure \ac{SYM} \SUTS
theory with dynamical scale given by $\Lambda^6_L=\Pf(S)\, \Lambda^3$. The
dynamics of the low-energy \ac{SYM} in turn generates a gaugino condensate
implying the existence of an effective superpotential
\begin{equation}
 \mathscr{W}_{\mathrm{dyn}}=\Lambda_L^3=
 \left(\Lambda^3\,\Pf (S) \right)^{\nicefrac{1}{2}}\,.
\end{equation}
It is easy to see that this superpotential stabilizes $S$ near the origin.

The main lesson we learn from this example is a possibility that the \ac{RG}
flow may change the chiral properties of the theory and, in particular, may change
the number of chiral generations. Here we define a chiral generation as a field
transforming in an antisymmetric representation of the chiral symmetry
accompanied by an appropriate number of fields in an antifundamental
representation as required by anomaly cancellation conditions. Then the net
number of generations is given by a difference between number of fields in an
antisymmetric representation and in a conjugate antisymmetric representation, 
\ytableausetup{boxsize=0.35em} 
$\nu = n_{\vphantom{\overline{\ydiagram{1,1}}}\ydiagram{1,1}} -
n_{\overline{\ydiagram{1,1}}}$. For example, in our example with $\SU6$ chiral
symmetry the number of generations is given by $n_{\rep{15}}-n_{\crep{15}}$.
This definition is chosen such that it can be used throughout this study, and
coincides with what one calls a generation in \SU5 \acp{GUT}. From the $\SU6$
perspective, our \ac{UV} model is a one-generation  model containing an
antisymmetric, $\rep{15}$, and two  antifundamental,  $\crep{6}$, of \SU6. On
the other hand, the \ac{IR} theory has no massless  chiral superfields even
while the chiral symmetry remains unbroken.

While the construction of \cite{Razamat:2020kyf} decreases the number of chiral
generations in the \ac{IR}, we will show in the following section that
non-perturbative dynamics may also lead to an increase in the number of chiral
generations.  As we will see, the existence of generation flow offers immense
opportunities for model building both in field theory (\Cref{sec:SU5_language})
and string theory (\Cref{sec:Explicit_string_models}).

\section{Generation flows in \acp{GUT}}
\label{sec:SU5_language}

The supersymmetric gapped fermion model reviewed in the previous section is
based on an \SUTS \SCONFG theory with \SU6 global symmetry.
Generalizations to \SCONFG $\SP{2N}$ with $\SU{2N+4}$ global symmetry are
straightforward \cite{Razamat:2020kyf}.\footnote{See also discussion in \cite{RSSW}.}
 However, for phenomenological purposes one is interested
in similar models with \SU5 or $\SU3\times \SU2\times \U1$ global symmetry which
can then be identified with the GUT or the \ac{SM} gauge group. As shown
in \cite{Razamat:2020kyf}, this can be easily achieved simply by considering the
model of \Cref{sec:gap} and identifying \ac{GUT} or \ac{SM} gauge group with
an appropriate subgroup of \SU6. 

For example, to construct a one-generation $\SU5\times \SUTS$ theory which
behaves as a pure SYM \SU5 in the \ac{IR}, one decomposes elementary fields of
the model under \SU5 as follows
\begin{equation}
\label{su5-embed}
S:(\rep{15},\rep{1}) \rightarrow T:(\rep{10},\rep{1})\oplus F:(\rep{5},\rep{1})\,, 
~~~ Q:(\crep{6},\rep{2})\rightarrow \overline{F}':(\crep{5},\rep{2})\oplus \phi:(\rep{1},\rep{2})\,.
\end{equation}
The tree-level superpotential~\eqref{eq:su2s-tree} and dynamical superpotential
\eqref{eq:sconfine} can be easily written in the \SU5 language. One can verify 
that the UV description corresponds to a one-generation model complemented
by a single vector-like flavor in a fundamental representation. As we learned
in \Cref{sec:SU6model}, the \SCONFG dynamics leads to a unique ground state
with an unbroken chiral symmetry and no light matter fields.

We are now ready to generalize the mass gap construction of \ac{RT}~\cite{Razamat:2020kyf} 
to obtain models where the number of chiral generations is changed through 
renormalization group flow but remains nonzero both in the UV and the IR. 
As we will see shortly, the RG flow may lead both to an increase and a
decrease in the effective number of chiral generations. The latter can be
achieved in two ways. In the first approach, as in the model of 
\Cref{sec:SU6model}, some of the chiral elementary fields acquire masses by
partnering with the chiral composites generated by confining dynamics. As a
result, all the massless degrees of freedom in the \ac{IR} are elementary fields
of the theory. Just like in the model of \Cref{sec:SU6model}, the chirally
symmetric vacuum is a unique ground state of this theory. The second approach is
reminiscent of the construction first introduced in \cite{Strassler:1995ia, Nelson:1996km}. 
In this approach, some of the massless fields in the \ac{IR} are composites even as 
other composites may become massive. Generically, models in this class retain the 
quantum moduli space and only one vacuum on this moduli space is chirally symmetric. 
Since \ac{IR} degrees of freedom, including the massless composites, are to be 
identified with the \ac{SM} multiplets, the motion along this moduli space is 
equivalent to motion along $D$-flat directions of a \ac{GUT} or the \ac{SM}. Note 
that the mechanism utilized in the second approach may also lead to an increase 
in the effective number of generations.

\subsection[4 -> 3 generation flow]{\FTT generation flow}

We can now detail our general observations by building an explicit model of 
downward generation flow. Let us start with a more straightforward example,
where  the number of chiral generations decreases in the IR while all the
composites are  heavy. In particular, we construct a \FTT model, i.e.\ a model
containing 4 generations in the UV and 3 generations in the \ac{IR}. The matter
fields of the  model and their quantum numbers are presented in
\Cref{tab:4to3a}. Note that this  matter content comprises the fields appearing
in \eqref{su5-embed} complemented  by three chiral flavors of \SU5 i.e.\ three
copies of $T\oplus\overline{F}$.  Thus, this is a four-generation model. It is
easy to see that \SUTS dynamics is  not affected by the introduction of
additional chiral multiplets as long as one  linear combination of the $T_i$'s
has the Yukawa coupling with $\overline{F}'$  and $\phi$ that is implied by the
superpotential \eqref{eq:su2s-tree}. Indeed, at low energies $\SU2_s$ charged
fields confine into $\overline{\Composite{T}}\sim \overline{F}^\prime
\overline{F}^\prime/\Lambda$ and $\overline{\Composite{F}}\sim
\overline{F}^\prime\phi/\Lambda$. The transformation properties of the IR
degrees of freedom are given in \Cref{tab:4to3b}. Finally, in the IR the
superpotential~\eqref{eq:su2s-tree} behaves like a mass term pairing composites
$\overline{\Composite{F}}$  and $\overline{\Composite{T}}$ with $F$ and one copy
of $T$, respectively. Repeating the analysis of \Cref{sec:gap} one concludes 
that the classical flat directions parameterized by $F$ and $T$ are stabilized
non-perturbatively.

\begin{table}[t!]
\centering
    \begin{subtable}[b]{0.4\linewidth}
    \centering
        \begin{tabular}{rcl}
          \toprule
          \# & irrep & label \\
          \midrule
           4 & $\left(\rep{10},\rep{1}\right)$ & $T$ \\
           2 & $\left(\crep{5},\rep{1}\right)$ & $\overline{F}$ \\
          \midrule
           1 & $\left(\crep{5},\rep{2}\right)$ & $\overline{F}^\prime$ \\
           1 & $\left(\rep{1},\rep{2}\right)$ & $\phi$ \\
           \midrule
           1 & $\left(\rep{5},\rep{1}\right)$ & $F$ \\
           1 & $\left(\crep{5},\rep{1}\right)$ & $\overline{F}$ \\
          \bottomrule
        \end{tabular}
        \caption{Unconfined spectrum.}
        \label{tab:4to3a}
    \end{subtable}
    \begin{subtable}[b]{0.4\linewidth}
    \centering
        \begin{tabular}{rcl}
            \toprule
                \# & irrep & label \\
            \midrule
               4 & $\left(\rep{10},\rep{1}\right)$ & $T$ \\
               4 & $\left(\crep{5},\rep{1}\right)$ & $\overline{F},\overline{\Composite{F}}$ \\
            \midrule
               1 & $\left(\crep{10},\rep{1}\right)$ & $\overline{\Composite{T}}$ \\
               1 & $\left(\rep{5},\rep{1}\right)$ & $F$ \\
            \bottomrule
        \end{tabular}
        \caption{Confined spectrum.}
        \label{tab:4to3b}
     \end{subtable}
\caption{Summary of the $\SU{5}\times \SU{2}_s$ quantum numbers of the chiral
superfield content of the \FTT model. The vector-like pair at the bottom of
\Cref{tab:4to3a} can be decoupled, resulting in a separate \FTT model.}
\label{tab:4to3}
\end{table}

\bigskip

Let us consider a generalization by noting that the symmetries of the model
allow  a mass term for the vector-like pair $F \oplus \overline F$. With this
mass term, the full \ac{UV} superpotential becomes
\begin{equation}
    \mathscr{W} = y_1\. T \overline F' \overline F' + y_2\. F\overline F'\phi +
	m\. F \overline F\,.
\end{equation}
Note that the additional mass term and $y_1 \neq y_2$ explicitly break the \SU6 
symmetry. Neither $F$ nor $\overline F$ are charged under \SUTS, thus the confined 
spectrum of the model (\Cref{tab:4to3b}) does not change. In the \ac{IR}, the 
superpotential becomes
\begin{equation}
    \mathscr{W}= \overline{\Composite{T}}\,
	\overline{\Composite{T}}\,\overline{\Composite{F}} + c_1\. 
	\Lambda\. T\. \overline{\Composite{T}} + 
	c_2\. \Lambda\. F\. \overline{\Composite{F}} + m\. F\. \overline{F}\,,
\end{equation}
where the first term is the \SCONFG superpotential \Cref{eq:sconfine}. A simple
analysis shows that in the presence of the mass term the model possesses a 
quantum moduli space satisfying the condition
\begin{equation}
    c_2 \Lambda\. \overline{\Composite{F}} + m\. \overline{F} = 0\,.
\end{equation}
While at a generic point on the moduli space the chiral \SU5 symmetry is broken, 
the \SCONFG vacuum where one generation acquires a mass survives at 
$\overline{\Composite{F}} = \overline F = 0$. This leaves three light generations, 
two made up entirely of elementary fields and another where the $\crep{5}$ is made 
up of a linear combination of $\overline{\Composite{F}}$ and $\overline F$. This 
lays out two interesting limits. In the limit $m\rightarrow 0$, the light generations 
are entirely composed of elementary fields, $\overline{\Composite{F}}=0$, and the 
chirally symmetric vacuum is stabilized as in \Cref{sec:gap}. We refer to this 
as the \ac{RT} limit because all composite fields decouple. In the limit 
$m\rightarrow \infty$, one of the three light generations has a composite 
$\crep{5}$. We refer to this limit as the \ac{NS} limit due to the appearance 
of light composite fields. At finite mass, there is a flat direction which can 
be parameterized by $\overline{\Composite{F}}$. For the purposes of phenomenology, 
$\overline{\Composite{F}}$ would play the role of a \ac{SM} multiplet; motion 
along the moduli space of this model corresponds to motion along $D$-flat 
directions of a \ac{GUT} (or the \ac{SM}).

\subsection[2 -> 3 generation flow]{\TTT generation flow}

The \ac{NS} limit of the model discussed above resulted in a theory with a
composite $\crep{5}$ while the number of $\rep{10}$'s (i.e. number of
generations) was smaller in the \ac{IR}. On the other hand, original models of
\cite{Strassler:1995ia,Nelson:1996km} had a composite $\rep{10}$ in the \ac{IR}
thus increasing the number of generations. That construction can be interpreted 
as an upward generation flow.  Let us discuss a variation of that model where
the  starting point of RG flow contains two chiral generations while the end
point in the \ac{IR} has three chiral generations, i.e.\ a \TTT model. 

Once again we consider a model with the symmetry group $\SU5\times \SU2_s$, whose
matter content and charges are given in \Cref{tab:2to3a}. The tree-level 
superpotential in terms of the \ac{UV} degrees of freedom is
\begin{equation}
    \mathscr W = y\. \overline{F}\. F^\prime \phi \,.
\end{equation}
When the non-perturbative dynamics is included, the \ac{IR} superpotential becomes
\begin{equation}
   \mathscr W = \Composite{T}\.\Composite{T}\.\Composite{F}
     + c\.\Lambda\. \overline{F}\.\Composite{F} \,,
\end{equation}
where $\Composite{T} \sim F' F'/\Lambda$  and $\Composite{F} \sim
F'\phi/\Lambda$.

\begin{table}[t!]
\centering
    \begin{subtable}[b]{0.4\linewidth}
    \centering
        \begin{tabular}{rcl}
          \toprule
          \# & irrep & label \\
          \midrule
           2 & $\left(\rep{10},\rep{1}\right)$ & $T$ \\
           4 & $\left(\crep{5},\rep{1}\right)$ & $\overline{F}$ \\
          \midrule
           1 & $\left(\rep{5},\rep{2}\right)$ & $F^\prime$ \\
           1 & $\left(\rep{1},\rep{2}\right)$ & $\phi$ \\
          \bottomrule
        \end{tabular}
        \caption{Unconfined spectrum.}
        \label{tab:2to3a}
    \end{subtable}
    \begin{subtable}[b]{0.4\linewidth}
    \centering
        \begin{tabular}{rcl}
            \toprule
                \# & irrep & label \\
            \midrule
               3 & $\left(\rep{10},\rep{1}\right)$ & $T,\Composite{T}$ \\
               3 & $\left(\crep{5},\rep{1}\right)$ & $\overline{F}$ \\
            \midrule
               1 & $\left(\crep{5},\rep{1}\right)$ & $\overline{F}$ \\
               1 & $\left(\rep{5},\rep{1}\right)$ & $\Composite{F}$ \\
            \bottomrule
        \end{tabular}
        \caption{Confined spectrum.}
        \label{tab:2to3b}
     \end{subtable}
\caption{Summary of the $\SU{5}\times \SU{2}_s$ quantum numbers of the chiral superfield content of the \TTT model.}
\label{tab:2to3}
\end{table}

It is convenient to analyze the behavior of this superpotential by going 
along a flat direction parameterized by $\overline F$. Without loss of 
generality we can assume that the VEV of $\overline F$ lives in a single 
component, say  $\overline F_5$. At large VEV, the global symmetry is broken 
from \SU{5} to \SU{4}, and one pair of doublets, the 
one corresponding to the $\Composite{F}_5$ meson, becomes heavy and can be 
integrated out. Along this flat direction the superpotential becomes
\begin{equation}
\label{eq:effective-deformed}
 \mathscr W=\Composite{F}_5 (\Pf' \Composite{T}+ \overline{F}_5)\,,
\end{equation}
where prime on the Pfaffian indicates that it is taken only over the light mesons
comprising a $\rep{6}$-plet of the remaining $\SU4$ symmetry.
Note that at this stage $\Composite{F}_5$ is not a dynamical field since it is a
meson made out of heavy doublets. 
At the same time, the $\overline{F}_5$ \ac{VEV} remains arbitrary albeit related 
to the $\Composite{T}$ \acp{VEV} by the $\Composite{F}_5$ equation of motion,
\begin{equation}
\label{eq:effective-LowE}
 \Pf' \Composite{T}+\overline F_5=0\,.
\end{equation}
Upon a careful inspection of~\eqref{eq:effective-deformed}
and~\eqref{eq:effective-LowE},  one notices that they correspond to the
superpotential and one of the equations  of motion of a four-doublet theory with
a deformed moduli space, a dynamical scale $\Lambda_L^{6}=\overline
F_5\Lambda^{5}$, and the meson $\Composite{F}_5$ playing a role of Lagrange
multiplier. We see that for each nonvanishing value of $\overline F_5$ the
effective theory possesses a quantum deformed moduli space, i.e.\ it exhibits  
confinement with chiral symmetry breaking. Furthermore, the scale of chiral
symmetry breaking is parameterized by $\overline F_5$. While the effective
description in terms of four-doublet theory is only valid at large $\overline
F_5$, the solution of the $\Composite{F}_5$  equation of motion is valid
everywhere on the quantum moduli space up to a  $\SU5$ symmetry transformation.
In particular, the chirally symmetric vacuum  $\Pf' \Composite{T}=\overline
F_5=0$ belongs to the quantum moduli space.

\bigskip

Note that the models introduced in this section differ in their  quantum moduli
spaces and their low-energy spectra. In the \ac{RT} limit of the \FTT model,
there is a unique, \SCONFG vacuum. All composite degrees of freedom become 
massive via the \ac{RT} mechanism, and there are three light generations made 
out of the elementary fields. In the \TTT model and the \ac{NS} limit of the
\FTT model,  there remains a quantum moduli space of vacua parameterized 
by the VEV of $\overline{\Composite{F}}$ (or equivalently $\overline F$),
respectively, which includes the chirally symmetric vacuum. In the \TTT model,
one of the three light generations contains a composite \rep{10},  while at
finite mass, the \FTT model has a \crep{5} which is partially composite and
partially elementary.

In the following sections, we will show how these models can arise naturally in 
string model building, providing examples of phenomenologically viable string
models  which would have previously been ruled out by the tree-level analysis of
the models.

\section{Generation flow in string models}
\label{sec:Explicit_string_models}

Given the possibility of generation flow discussed in
\Cref{sec:SU6model,sec:SU5_language}, we will now turn our attention to string
model building. Why can generation flow be relevant for string models? In string
phenomenology, one tries to connect string theory to the real world (cf.\ e.g.\
\cite{Ibanez:2012zz}). In practice, this often amounts to searching for a string
compactification which reproduces the \ac{SM} in its low-energy limit. When
constructing a string model, one chooses a framework, such as one of the
perturbative string theories, and compactifies it down to four dimensions. The
step of compactification consists of making an assumption on the geometry of
compact dimensions  (in principle one also must show that the emerging setup is
stable, i.e.\ string moduli describing the size and shape of compact space are
stabilized). However, attempts to build realistic models often fail already at
an earlier stage because the zero-modes do not comprise the \ac{SM} matter. This
could mean that one has chiral exotics, or just not the right number of
generations. It is the latter possibility where generation flow, as discussed in
\Cref{sec:SU5_language}, can be important.\footnote{It is conceivable that more
generally chiral exotics can be removed along the lines of \Cref{sec:SU6model}
(cf.\ \cite{Dimopoulos:1996dz} for an example). It will be interesting to work
out the detailed conditions for this to happen.} In practice, when determining
the number of generations, one looks at the tree-level predictions. However, as
discussed in \Cref{sec:SU6model,sec:SU5_language}, the number of generations
obtained this way may differ from the true number of chiral generations in the
low-energy effective theory.\footnote{It is known that chirality-changing
phase transitions can occur in string compactifications
\cite{Kachru:1997rs,Douglas:2004yv,Anderson:2015cqy}. In this work we focus on
generation flow that can be understood in terms of field-theoretic
supersymmetric gauge dynamics with an \SCONFG \SUTS as in
\Cref{sec:SU6model,sec:SU5_language}. It will be interesting to see whether
there is a deeper relation between these phenomena.} It is therefore interesting to study the
question to which extent models of the type discussed earlier can be obtained
from string theory.

It is not the purpose of the present paper to construct a fully realistic model
exhibiting generation flow. Rather, we will collect evidence for the existence
of such models. To keep our discussion simple, and in order to relate our
findings to \Cref{sec:SU5_language}, we will look for \SU5 models rather than
models with \ac{SM} gauge group. However,  we expect that the results carry over to models with  the \ac{SM} gauge
group after compactification.

\subsection{Model scan}

In what follows, we focus on orbifold compactifications of the ($\E8\x\E8'$)
heterotic string~\cite{Dixon:1985jw,Dixon:1986jc}, which can be efficiently
constructed with the \texttt{orbifolder}~\cite{Nilles:2011aj}. We will  collect
evidence for the existence of globally consistent string compactifications that
have either two or four generations of \ac{SM} matter at tree level, but in fact
have three generations in their low-energy effective description.  That is, we
will present evidence for the existence of stringy versions of the \FTT and \TTT
models discussed in \Cref{sec:SU5_language}. 

The \texttt{orbifolder} allows us to compute a 4D model from certain input data,
which comprises the geometry of the orbifold and the so-called gauge embedding.
The latter essentially describes how the geometric operations of the 6D
space-like compact dimensions act on the $\E8\x\E8'$ lattice. This
determines not only what the residual gauge symmetry of the model is but
also the spectrum. In more detail, the \texttt{orbifolder} provides us with the
continuous and discrete gauge symmetries after compactification as well as the
chiral spectrum of the model.

By using the \texttt{orbifolder}, we obtained a large sample of supersymmetric 
heterotic orbifold models with the following properties: 
\begin{itemize}
 \item orbifold geometry $\Z2\times\Z4$ (1,1) (see \cite{Fischer:2012qj} for the
	notation, and~\cite{MayorgaPena:2012ifg} for details of the geometry);
 \item 4D gauge group $\mathcal{G}_\text{4D} \supset \SU5\x\SU2_s$ (where we
 labeled the second factor ``$s$'' to indicate that this \SU2 plays the same
 role as in our earlier discussion in \Cref{sec:SU6model,sec:SU5_language});
 \item the \SU5 and $\SU2_s$ gauge groups emerge each from a different \E8
	   factor of the original heterotic string;
 \item a net number of $n$ \SU5 GUT generations, with no representation
       $(\rep{10},\rep2)$ \emph least one representation
       $(\crep5,\rep2)$ or $(\rep5,\rep2)$;
 \item at least one ``flavon'' field transforming as $(\rep1,\rep2)$; other
       fields of this type could in principle be decoupled from low energies; 
 \item a (large) number of $\SU5\x\SU2_s$ singlets;
 \item additional non-Abelian gauge factors under which the \SU5 charged fields are singlets; and
 \item additional \U1 factors which can be broken along $D$-flat directions 
       without breaking $\SU5\times\SU2_s$.
\end{itemize}
Our scan yielded several models in which \SCONF can change the
number of chiral representations. 

\subsection{Models}

Rather than providing the reader with an extensive survey, we focus on
two sample models defined in the Appendix. In more detail, we discuss
\begin{itemize}
 \item a \FTT model (cf.\ \Cref{tab:model203}) in which the $4^\mathrm{th}$
  chiral generation acquires a mass and decouples through,
  and
 \item a \TTT model (cf.\ \Cref{tab:model44}) in which the  $3^\mathrm{rd}$
  chiral generation emerges from states that are vector-like under
  \SU5 through a variant of the \ac{RT} effect, in which a chiral
  $\rep{10}\oplus\crep{5}$ arises as a composite of 
  $(\rep{5},\rep{2})\oplus(\rep{1},\rep{2})\oplus2(\crep{5},\rep{1})$
\end{itemize}
Both models have the virtue that the \SU5 and $\SU2_s$ factors come from different
\E8's. Consequently, $\SU2_s$ can naturally be more strongly coupled than \SU5
(cf.\ e.g.\ \cite{Ibanez:1986xy}).

\begin{table}[t!]
\centering
 \begin{subtable}[t]{0.4\linewidth}
    \centering
    \begin{tabular}{rcl}
	  \multicolumn{3}{c}{\FTT model}\\
      \toprule
      \# & irrep & label \\
      \midrule
       4 & $\left(\rep{10},\rep{1}\right)$ & $ T$\\
       4 & $\left(\crep{5},\rep{1}\right)$ & $ F$\\       
      \midrule
       7 & $\left(\crep{5},\rep{1}\right)$ & $ F$\\
       9 & $\left(\rep{5},\rep{1}\right)$  & $ \overline F$\\       
       1 & $\left(\crep{5},\rep{2}\right)$ & $ \overline F'$\\       
      \midrule
     170 & $\left(\rep{1},\rep{1}\right)$ & $ N$\\
      27 & $\left(\rep{1},\rep{2}\right)$ & $ \phi$\\
      \bottomrule
    \end{tabular}
    \caption{The first block contains four chiral generations of \SU5 matter.}
    \label{tab:model203}
 \end{subtable}
 \hspace{0.1\linewidth}%
 \begin{subtable}[t]{0.4\linewidth}
    \centering
    \begin{tabular}{rcl}
	  \multicolumn{3}{c}{\TTT model}\\
      \toprule
      \# & irrep & label \\
      \midrule
       2 & $\left(\rep{10},\rep{1}\right)$ & $T$\\
       2 & $\left(\crep{5},\rep{1}\right)$ & $F$\\
      \midrule
      10 & $\left(\crep{5},\rep{1}\right)$ & $F$\\       
       8 & $\left(\rep{5},\rep{1}\right)$  & $\overline F$\\
       1 & $\left(\rep{5},\rep{2}\right)$  & $F'$\\      
      \midrule
     240 & $\left(\rep{1},\rep{1}\right)$  & $N$\\
      41 & $\left(\rep{1},\rep{2}\right)$  & $\phi$\\
      \bottomrule
    \end{tabular}
    \caption{The first block represents two chiral families of an \SU5 GUT.}
    \label{tab:model44}
 \end{subtable}
\caption{Summary of the $\SU5\times\SU2_s$ quantum numbers of the (left-chiral) 
    massless matter spectra of heterotic orbifold models with (a) \FTT and 
    (b) \TTT \SU5 generation flow. These models have (a) four and (b) two chiral 
	generations at tree level, respectively, but three chiral generations in the 
	low-energy effective description due to $\SU2_s$ strong dynamics.
	The second (third) block of each table consists of states that 
	are vector-like (invariant) under \SU5.} 
\end{table}

\subsubsection*{A stringy \FTT model}

The model defined by the parameters provided in~\Cref{eqs:shifts_WL_4-1} results
in the 4D gauge group $\mathcal{G}_\text{4D} = \SU5\x\SU2_s\x[\SU2^5\x\U1^6]$. 
The gauge factors in the brackets can be broken along
$D$-flat directions. Since the Lagrange density is invariant under
complexified gauge transformation, we can infer that nontrivial solutions to the
$F$-term equations preserve supersymmetry \cite{Buccella:1982nx,Luty:1995sd}.
We are then left with $\mathcal{G}_\mathrm{unbroken}=\SU5\x\SU2_s$. 

Before discussing the \FTT properties of this model, let us comment on the
possibility to break $\SU2_s$ along $D$-flat directions. In this case, we will
obtain a vacuum with 4 generations of an \SU5 GUT, i.e.\ 4 copies of
$\rep{10}\oplus\crep{5}$ while the other states are now vector-like and pick up
masses proportional to the \acp{VEV} of the \SU5 singlets that got switched on.
According to the usual string phenomenology practices, we would thus label this
model an unrealistic 4-generation model, not worth being considered further.

On the other hand, if we leave $\SU2_s$ unbroken, in a generic vacuum we obtain
in an intermediate step a model with 4 copies of
$\left(\rep{10},\rep{1}\right)$,  2 copies of $\left(\crep{5},\rep{1}\right)$, a
$\left(\crep{5},\rep{2}\right)$ and a  $\left(\rep{1},\rep{2}\right)$. Since
string selection rules do not forbid the  corresponding couplings, the other
states of \Cref{tab:model203} acquire masses  proportional to the \acp{VEV} of
the $\SU5\times\SU2_s$ singlets. Conceivably, there also exist special
string vacua that can allow for an extra massless vector-like  pair
$(\rep5,\rep1)\oplus(\crep5,\rep1)$. This brings us to either of the \FTT models
discussed in \Cref{sec:SU5_language}, and summarized in \Cref{tab:4to3a}.  As we
have seen there, due to the $\SU2_s$ strong dynamics,
$\left(\crep{5},\rep{2}\right)$  and $\left(\rep{1},\rep{2}\right)$ condense
together to build a $\crep{5}$ and  condensates of
$\left(\crep{5},\rep{2}\right)$ yield an \SU5 antigeneration $\crep{10}$. Since
there are no string selection rules prohibiting the couplings, we thus expect
this antigeneration to pair up with a linear combination of the 4 generations,
and we are left with a 3-generation model at low energies.

An important condition for the strong $\SU2_s$ dynamics to play out as described
is that $\SU2_s$ is much more strongly coupled than \SU5. Since these two gauge
factors originate from different \E8's, it is plausible that this happens
\cite{Ibanez:1986xy,Dixon:1990pc,Stieberger:1998yi}. However, a detailed
computation of the string thresholds is beyond the scope of this study.

\subsubsection*{A stringy \TTT model}
\label{sec:model2+1}

The model defined by the parameters provided in~\Cref{eqs:shifts_WL_2+1}
results in the 4D gauge group $\mathcal{G}_\text{4D} =
\SU5\x\SU2_s\x[\SU2^2\x\U1^9]$. As in the previous model, the gauge factors in
parentheses can be spontaneously broken along $D$-flat directions while preserving
supersymmetry. The corresponding massless spectrum after compactification is
summarized in~\Cref{tab:model44}, where we only display the quantum numbers with
respect to $\SU5\x\SU2_s$. After switching on the \acp{VEV} of $\SU5\x\SU2_s$
singlets, we are left with 2 copies of $(\rep{10},\rep{1})$, 4 copies of
$(\crep{5},\rep{1})$, and 1 instance of $(\rep{5},\rep{2})$ and $(\rep{1},\rep{2})$, 
reproducing the spectrum of the \TTT model presented in \Cref{tab:2to3a}.

If we also break $\SU2_s$ along $D$-flat directions, we obtain a vacuum with an
\SU5 \ac{GUT} symmetry and two generations of $\rep{10}\oplus\crep{5}$. In the
traditional approach, we would thus label the model as an unrealistic
2-generation model that is to be discarded.

However, this conclusion changes if we look at vacua where $\SU2_s$ confines. In this case, according to our discussion of the \TTT model in
\Cref{sec:SU5_language}, we can obtain a third generation from $\SU2_s$ strong
dynamics. In particular, the $(\rep5,\rep2)$ builds a condensate that 
behaves as the $\rep{10}$-plet of a third generation of an \SU5 \ac{GUT}.
This means that this model admits 3-generation vacua and cannot be
ruled out immediately.

\subsection{Discussion}

The examples discussed in this section represent evidence for the existence of 
globally consistent string models with generation flow. In order to keep the discussion 
simple, we have focused on \SU5 models. However, we expect that qualitatively similar 
models with the \ac{SM} gauge symmetry and matter content at low energies exist. We have 
verified that one can break extra gauge factors and decouple exotics by switching 
on \acp{VEV} along $D$-flat directions. We are thus guaranteed~\cite{Buccella:1982nx,Luty:1995sd} 
that there are supersymmetric configurations that have the features we describe. 
While we did verify that there are no symmetries prohibiting the required couplings, we did 
not compute their coefficients, nor did we explicitly verify that all
directions/moduli are stabilized.

Our findings lead to the following picture. In string models, one can readily
count the net number of generations at the tree-level. However, some models may 
have vacua where the true number of
chiral generations differs from the tree-level prediction. This means that
model scans in the past may have missed interesting, possibly realistic models.
It will be interesting to study such constructions in more detail. 

As a side remark, let us note that the matter content as well as the gauge and
continuous symmetries of the \ac{RT}-like model discussed in
\Cref{sec:SU6model} fit into a \rep{27}-plet of
$\text{E}_6$. This is evident from the branching (cf.\ e.g.\
\cite{Slansky:1981yr})
\begin{subequations}
\begin{align}
 \text{E}_6&\to\SU6\times\SU2_s\\
 &\to\SU5\times\SU2_s\times\U1\;,\\
 \rep{27}&\to(\crep{6},\rep{2})\oplus(\rep{15},\rep{1})\\
 &\to(\crep{5},\rep{2})_1\oplus(\rep{1},\rep{2})_{-5}\oplus
 (\rep{10},\rep{1})_{-2}\oplus(\rep{5},\rep{1})_4\;.
\end{align}
\end{subequations}
That is, while the representation content of the model may at first sight look a
bit peculiar, it turns out to fit in a single chiral representation of an
exceptional group. In fact, \E6 is the only exceptional group admitting
complex representations, and the \rep{27}-plet is its smallest representation.
From this perspective it is not too surprising that variants of this model can
be obtained from string theory. Note, however, that in the models which we
presented, \SU5 and $\SU2_s$ stem from different \E8 groups, which favors the
possibility that $\SU2_s$ becomes strongly coupled while \SU5 does not.

\section{Summary}
\label{sec:Summary}

We have studied the effects of non-perturbative $s$-confining dynamics on the 
effective number of chiral generations in supersymmetric models of particle 
physics.
We emphasized that this number can flow either upward or downward 
because confinement may result in the appearance of chiral composites. In turn,
these composites may either serve as new light chiral generations or lift
existing chiral generations by partnering with other chiral fields in mass terms.
We referred to these phenomena as generation flow.

Our focus was on \FTT and \TTT generation flow, such that in the \ac{IR} there 
are three generations of (a \ac{GUT} completion of) the \ac{SM}. We analyzed 
the non-perturbative dynamics and verified that in our models the 
\SCONFG vacuum is not destabilized by the non-perturbative dynamics driving 
the generation flow. We stress 
that this conclusion is model dependent.

As we have shown, there is strong evidence that generation flow arises in 
globally consistent string compactifications. In particular, we have constructed
explicit \FTT and \TTT models resulting from orbifold compactifications of the 
heterotic string. Therefore, more care than previously appreciated has to be
taken when scanning for realistic string models. There can be models which
appear to yield an unrealistic number of generations but are saved by generation
flow. Furthermore, the strong dynamics that reduces the number of generations
may be exploited to decouple chiral exotics of string  models. Hence, the
phenomenological viability of string compactifications with such exotics should
be further investigated.

\subsection*{Acknowledgments}

We would like to thank Eric Sharpe for pointing out
\cite{Kachru:1997rs,Douglas:2004yv} to us.
The work of M.R., Y.S., S.S.\ and M.W.\ was supported by the National Science
Foundation, under Grant No.\ PHY-1915005.  The work of S.R.-S., M.R.\ and S.S.\
was supported by UC-MEXUS-CONACyT  grant No.\ CN-20-38. This work was
performed in part at Aspen Center for Physics, which is supported by National
Science Foundation grant PHY-1607611.

\newpage
\appendix
\section{Orbifold model definitions}
\label{app:StringParameters}

In the bosonic formulation, a $\Z2\x\Z4$ (1,1) heterotic orbifold
compactification is defined by the shifts $V_1$ and $V_2$ of order 2 and 4,
respectively, as well as six discrete Wilson lines $W_a$, $a=1,\ldots,6$ of
order 2. These Wilson lines are restricted to satisfy $W_1=W_2$ and $W_5=W_6$ to
be compatible with the $\Z2\x\Z4$ point group of the
compactification.\footnote{See
e.g.~\cite{Bailin:1999nk,Ramos-Sanchez:2008nwx,Vaudrevange:2008sm}  for reviews
on orbifold compactifications, and~\cite[Section 4]{MayorgaPena:2012ifg} for more details on 
this specific orbifold geometry.} These parameters can be used as input in the 
\texttt{orbifolder}~\cite{Nilles:2011aj} to obtain the corresponding massless
spectrum and compute the superpotential of the associated low-energy effective
field theory.

\subsection[Details of the 4 -> 3 heterotic orbifold model]{Details of the \FTT heterotic orbifold model}

One heterotic orbifold model with geometry $\Z2\x\Z4$ (1,1) which yields
\FTT generations via the \ac{RT} scheme is defined by the following shifts and
Wilson lines (with $W_4=0$):
\begin{subequations}
\label{eqs:shifts_WL_4-1}
\begin{align}
  V_1 &= \left(-\tfrac{7}{4}, -\tfrac{1}{4}, -\tfrac{1}{4}, -\tfrac{1}{4}, -\tfrac{1}{4}, -\tfrac{1}{4}, \tfrac{1}{4}, \tfrac{7}{4}\right),  \left(    0,     0,     0,     0,     0,     0,     0,     0\right)\,,\\
  V_2 &= \left(\tfrac{3}{8}, \tfrac{1}{8}, \tfrac{1}{8}, \tfrac{1}{8}, \tfrac{3}{8}, \tfrac{9}{8}, -\tfrac{3}{8}, -\tfrac{3}{8}\right),  \left(   -1,     0,     0,     0, \tfrac{1}{4}, \tfrac{1}{4}, \tfrac{1}{4}, \tfrac{3}{4}\right)\,,\\
  W_1 & = W_2 = \left(    0,     0,     0,     0,     0,     0,     0,     0\right),  \left(   -1,    -1, \tfrac{1}{2}, \tfrac{3}{2}, -\tfrac{1}{2},     0, \tfrac{1}{2},     0\right)\,,\\
  W_{3} & = \left(    0,     0,     0,     0,     0,     0,     0,     0\right),  \left(-\tfrac{5}{4}, -\tfrac{5}{4}, \tfrac{1}{4}, \tfrac{3}{4}, \tfrac{3}{4}, \tfrac{7}{4}, -\tfrac{3}{4}, \tfrac{7}{4}\right)\,,\\
  W_5 & = W_6 = \left(   -1,    -1,     0,     1, \tfrac{3}{2}, \tfrac{1}{2}, \tfrac{1}{2}, \tfrac{3}{2}\right),  \left(    0,     0,     0,     0,     0,     0,     0,     0\right)\,.
\end{align}
\end{subequations}
The effective massless matter spectrum before decoupling of vector-like
representations and $\SU2_s$ confinement, obtained by the \texttt{orbifolder} 
is summarized in \Cref{tab:model203}.

\subsection[Details of the 2 -> 3 heterotic orbifold model]{Details of the \TTT heterotic orbifold model}

The orbifold parameters that define the $\Z2\x\Z4$ (1,1) heterotic orbifold
model  presented in section~\ref{sec:model2+1} are
\begin{subequations}
\label{eqs:shifts_WL_2+1}
\begin{align}
  V_1 &= \left(-\tfrac{1}{4}, -\tfrac{1}{4}, \tfrac{1}{4}, \tfrac{1}{4}, \tfrac{1}{4}, \tfrac{1}{4}, \tfrac{1}{4}, \tfrac{9}{4}\right),  \left(    0,     0,     0,     0,     0,     0,     0,     2\right)\,,\\
  V_2 &= \left(\tfrac{1}{8}, \tfrac{9}{8}, -\tfrac{7}{8}, -\tfrac{1}{8}, -\tfrac{1}{8}, -\tfrac{1}{8}, \tfrac{9}{8}, \tfrac{7}{8}\right),  \left(-\tfrac{1}{2},     0,     0,     0, \tfrac{1}{4}, \tfrac{1}{4}, \tfrac{3}{4}, -\tfrac{3}{4}\right)\,,\\
  W_{1} & = W_2 = \left(    1,     0,    -2,    -1,     0,     1,    -1,    -2\right),  \left(\tfrac{1}{4}, -\tfrac{3}{4}, -\tfrac{1}{4}, \tfrac{7}{4}, -\tfrac{3}{4}, \tfrac{3}{4}, -\tfrac{5}{4}, \tfrac{5}{4}\right)\,,\\
  W_{3} & = \left(-\tfrac{5}{4}, \tfrac{5}{4}, \tfrac{5}{4}, -\tfrac{7}{4}, -\tfrac{5}{4}, -\tfrac{5}{4}, \tfrac{1}{4}, -\tfrac{5}{4}\right),  \left(\tfrac{7}{4}, \tfrac{5}{4}, \tfrac{7}{4}, \tfrac{7}{4}, \tfrac{5}{4}, \tfrac{9}{4}, -\tfrac{1}{4}, \tfrac{9}{4}\right)\,,\\
  W_{5} & =  W_6 = \left(   -2, -\tfrac{1}{2},     0,     1, -\tfrac{1}{2},     1, \tfrac{1}{2}, \tfrac{3}{2}\right),  \left(-\tfrac{7}{4}, -\tfrac{1}{4}, -\tfrac{5}{4}, -\tfrac{5}{4}, \tfrac{7}{4}, \tfrac{1}{4}, -\tfrac{3}{4}, -\tfrac{7}{4}\right)\,,
\end{align}
\end{subequations}
and $W_4=0$.
Using these parameters as input of the \texttt{orbifolder}, one finds the
massless matter spectrum before decoupling of vector-like representations
and $\SU2_s$ confinement shown in \Cref{tab:model44}.

\bibliographystyle{NewArXiv}
\bibliography{GappedFermionsStrings}

\end{document}